# Magnetic Properties of One-Dimensional Quantum Spin System $Rb_2Cu_2Mo_3O_{12}$ Studied by Muon Spin Relaxation

Seiko OHIRA-KAWAMURA[1*], Keisuke TOMIYASU[2], Akihiro KODA[3], Dita P. SARI[4,5], Retno ASIH[4,5], Sungwon YOON[5,6], Isao WATANABE[4,5,6,7], and Kenji NAKAJIMA[1]

[1]*J-PARC Center, Japan Atomic Energy Agency, Tokai, Ibaraki 319-1195, Japan*
[2]*Department of Physics, Tohoku University, Sendai 980-8578, Japan*
[3]*Institute of Materials Structure Science, High Energy Accelerator Research Organization (KEK), Tokai, Ibaraki 319-1106, Japan*
[4]*Department of Physics, Osaka University, Toyonaka, Osaka 560-0043, Japan*
[5]*RIKEN Nishina Center, Wako, Saitama 351-0198, Japan*
[6]*Department of Physics, The Catholic University of Korea, Bucheon 14662, Korea*
[7]*Department of Condensed Matter Physics, Hokkaido University, Sapporo 060-0810, Japan*

*E-mail: seiko.kawamura@j-parc.jp

Magnetic properties of a one-dimensional frustrated quantum spin system $Rb_2Cu_2Mo_3O_{12}$ is investigated by the muon spin relaxation (μSR) technique. Although this system shows an incommensurate spin correlation in neutron scattering, it has not yet been determined whether the ground state reaches to a nonmagnetic spin-singlet state or not. In this study, our zero-field (ZF) μSR result undoubtedly reveals a nonmagnetic ground state, which establishes formation of the nonmagnetic incommensurate spin-singlet state in combination of the previous neutron scattering study. Furthermore, we found that the internal field is slightly enhanced below ~7 K. This temperature dependence resembles that of electric polarization under the magnetic field reported recently, indicating a possibility that a small change occurs even under ZF. In longitudinal-field measurements, a fast fluctuation around a narrowing region is observed. This relaxation is expected to reflect characteristics of formation of the spin-singlet state.

## 1. Introduction

Quantum spin system with spin frustration has been studied intensively, since they often show exotic spin state and dynamical behavior. The low-dimensional spin system where the first and second nearest neighbor exchange interactions, $J_1$ and $J_2$, are competing, such as one-dimensional (1D) zig-zag chain and 2D square lattice systems, is an especially interesting subject in such frustrated spin systems. For these systems, a ratio $\alpha \equiv J_2/J_1$ governs their variety of the magnetic ground states.

$Rb_2Cu_2Mo_3O_{12}$ is a 1D quantum spin system having a zig-zag chain structure of $Cu^{2+}$ ions with $S = 1/2$. There exists competition/frustration due to the nearest neighbor ferromagnetic and the second nearest neighbor antiferromagnetic exchange interactions. The interactions are estimated to be $J_1 = -138$ K and $J_2 = 51$ K, respectively, and thus $\alpha =$

-0.37 [1]. In the case where $J_1$ is ferromagnetic and $\alpha < -0.25$, the magnetic ground state is expected to be an incommensurate spin-singlet state [2]. In fact, a previous inelastic neutron scattering study observes the dynamical short-range incommensurate spin correlations [3]. It is also noted that ferroelectricity appears below $T_c \sim 8$ K under the magnetic field without magnetic transition [4]. However, there is no direct experimental evidence for the formation of a nonmagnetic ground state. Thus, we have studied the magnetic ground state and dynamics in this material by means of muon spin relaxation (μSR). The present study clearly reveals that the magnetic ground state of this system is nonmagnetic.

## 2. Experimental

Pellets of $Rb_2Cu_2Mo_3O_{12}$ were prepared after characterization by the magnetic susceptibility measurement. The zero-field (ZF) and longitudinal-field (LF) μSR experiments were carried out at two spectrometers, DΩ1 installed at the Materials and Life Science Experimental Facility (MLF), in Japan Proton Accelerator Research Complex (J-PARC) and ARGUS installed at the RIKEN-RAL Muon Facility, in Rutherford Appleton Laboratory. For the experiment at DΩ1, the pellet sample was wrapped with an aluminum foil to fix onto a $^4$He Mini Cryostat and was cooled down to 3 K. The LF up to 100 G was applied. At ARGUS, the measurements were performed at temperatures ranging from 0.3 to 50 K and from 0.06 to 0.6 K by using a bottom-loading $^3$He cryostat and a $^3$He-$^4$He dilution refrigerator, respectively. The pellet sample was directly mounted onto a sample holder of the cryostat with the Apiezon-N grease. The LF up to 3950 G was applied for the measurements with the $^3$He cryostat, while only ZF measurement was performed using the dilution refrigerator.

## 3. Results and Discussion

Figures 1(a) and 1(b) show the ZF-μSR time spectra observed at temperatures above 3 K with a Mini Cryostat and those above 0.3 K with a $^3$He cryostat, respectively. Slow muon spin relaxation due to a small nuclear dipole field is observed in the whole measured temperature region. The time spectra show no remarkable change in the temperature range from 20 to 150 K, while its relaxation becomes slow at higher temperatures due to a motional narrowing effect. The muon spin relaxation is gradually enhanced below 20 K, and then it becomes constant below ~1 K again. The enhancement of the muon spin relaxation is small and quite different from that observed for a magnetic transition or its precursor. This result clearly indicates that the magnetic ground state of this system is nonmagnetic.

The time spectra are fitted with a function

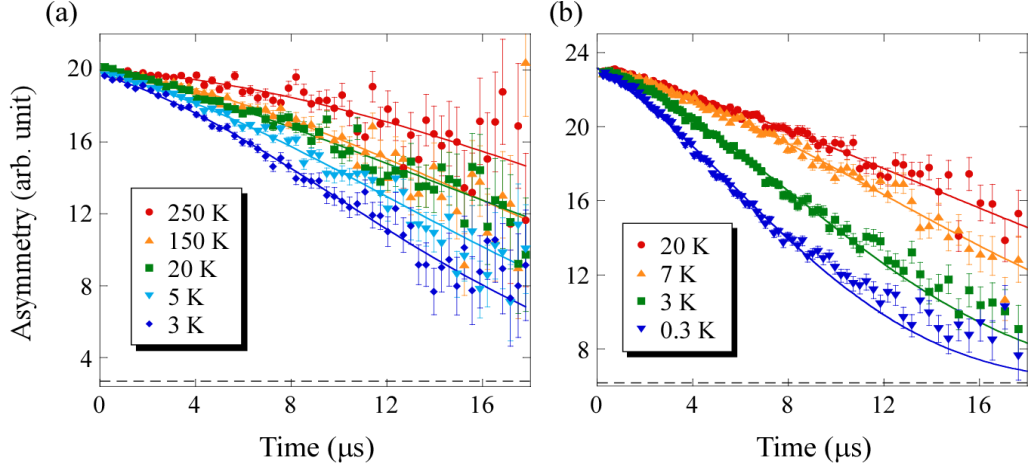

**Fig. 1.** ZF-µSR time spectra of Rb$_2$Cu$_2$Mo$_3$O$_{12}$ observed at temperatures (a) with a Mini Cryostat and (b) with a $^3$He cryostat. Baselines estimated from analyses are indicated by dashed lines.

$$A(t) = A\, G_{\mathrm{KT}}(\varDelta, t)\mathrm{e}^{-\lambda t}, \qquad (1)$$

where $A$ is the initial asymmetry, $G_{\mathrm{KT}}$ the static Gaussian Kubo-Toyabe function [5], $\varDelta$ the distribution width of the internal field, and $\lambda$ the muon spin relaxation rate. The temperature dependences of $\lambda$ and $\varDelta$ are shown in Figs. 2(a) and 2(b), respectively. It is noted that the time spectra above 150 K are explained as motional narrowing by introducing a dynamical Gaussian Kubo-Toyabe function (not plotted here) [5]. $\lambda$ is independent of temperature above 5 K, while it increases below 3 K with decreasing temperature and then reaches a constant value below ~1 K. The increase in $\lambda$ is very small in sharp contrast to that for strong suppression of spin fluctuation. The origin of the slight enhancement of the muon spin relaxation is discussed later with the LF-µSR results. $\varDelta$ exhibits a slight increase below ~7 K. Since a finite $\varDelta$ value originates from a random and static nuclear dipole field in usual cases, the value should be independent of temperature in the measured temperature range. However, in this system, a common value cannot reproduce the time spectra in the whole measured temperature region, even though we apply any conceivable functions. The increase in $\Delta$ is very small, which is similar in the order of the nuclear dipoles, and it does not indicate a magnetic order. We notice here that the temperature dependence of $\Delta$ resembles that of the electric polarization obtained in the magnetic field [4]. The ferroelectricity of this system is not detected in ZF and induced by the external magnetic field without magnetic order. It is theoretically predicted that noncollinear magnetism leads to the occurrence of electric polarization [6], though such a magnetic order is not detected in bulk measurements. As a possibility for the increase in $\Delta$, some slight change in a local magnetic state related to the field-induced ferroelectricity might be detected even under ZF.

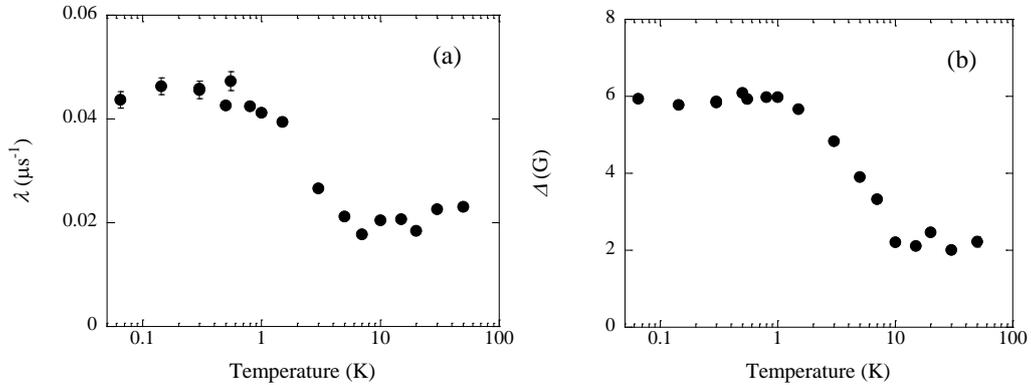

**Fig. 2.** Temperature dependences of (a) the relaxation rate $\lambda$ and (b) the distribution width of the internal field $\varDelta$.

The dynamical behavior observed in ZF-µSR can be investigated more in detail by LF decoupling measurements. The LF-µSR time spectra are reproduced by a product of an LF Kubo-Toyabe function $G_{KT}(\Delta, H_{LF}, t)$ and an exponential one for LF at and lower than 20 G and by a simple exponential function for LF higher than 20 G. Figures 3(a)-3(e) show the LF dependence of the relaxation rate $\lambda$ obtained at various temperatures. It is found there that the dynamical relaxation component contains two types of the fluctuations: $\lambda$ for the first component decreases with increasing LF and the second one is independent of LF up to high fields. Fitting $\lambda$ with a Redfield's model, the first (LF-dependent) component corresponds to the fluctuating internal field $H_{int} \sim 1$ G with a correlation time $\tau_c < 1$ µs. As both $H_{int}$ and $\tau_c$ are almost independent of temperature, it could not be intrinsic magnetism of this system. On the other hand, the second component is roughly independent of LF indicating a fast fluctuation around a motional narrowing region. In this region, we cannot obtain $H_{int}$ and $\tau_c$ individually, but obtain $\lambda = 2\gamma_\mu^2 H_{int}^2 \tau_c$, where $\gamma_\mu$ is the muon gyromagnetic ratio. The temperature dependence of $\lambda$ obtained by the fitting is shown in Fig. 3(f). This component shows some temperature dependence in contrast to the first component. $\lambda$ is almost constant in a paramagnetic region and drops around 10 K, which indicates formation of the spin-singlet state. Then $\lambda$ shows an upturn below 10 K. The increase in $\lambda$ in the nonmagnetic state is often observed in several spin-singlet systems such as spin-Peierls and charge separation ones [7]. It may be explained as an activation due to an energy gap or dilute and quasistatic free spins existing in the spin-singlet state. Therefore, the temperature dependence of this fluctuation component is expected to reflect characteristics of formation of the spin-singlet state or coexisting free spins with a small fraction.

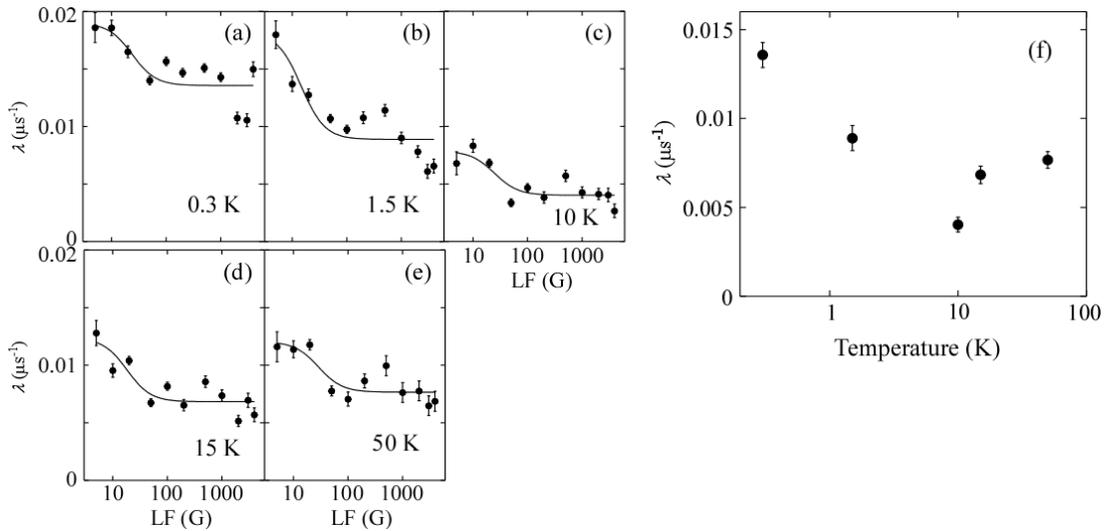

**Fig. 3.** LF dependence of $\lambda$ at (a) 0.3, (b) 1.5, (c) 10, (d) 15 and (e) 50 K. Solid lines are fits to a Redfield's equation described in the text. (f) Temperature dependence of the relaxation rate $\lambda$ for the LF-independent component obtained by the fitting.

**4.  Summary**

We investigated the magnetic ground state and dynamical properties of a 1D quantum spin system $Rb_2Cu_2Mo_3O_{12}$ by μSR. No remarkable change in the ZF-μSR time spectra was observed down to 0.06 K, which clearly indicates the nonmagnetic ground state. A slight increase in $\Delta$ below ~7 K may reflect some change in a local magnetic state related to the field-induced ferroelectricity, though any magnetic order is not observed in bulk measurements under the magnetic field. In the LF-μSR study, we found that the relaxation component contains two dynamical fluctuation. One of them, the fast fluctuation component around a motional narrowing region is expected to reflect characteristics of formation of the spin-singlet state or dilute free spins existing in the nonmagnetic phase. The present study reveals that the ground state in this system is incommensurate spin-singlet in combination of the previous neutron scattering study.

**Acknowledgment**

The μSR experiments performed at DΩ1 in J-PARC MLF and ARGUS in the RIKEN-RAL Muon Facility were based on the approved proposals, No. 2012B0124 and R494, respectively.